\documentclass[conference]{IEEEtran}
\IEEEoverridecommandlockouts
\usepackage{cite}
\usepackage{amsmath,amssymb,amsfonts}
\usepackage{algorithmic}
\usepackage{graphicx}
\usepackage{textcomp}
\usepackage{xcolor}
\def\BibTeX{{\rm B\kern-.05em{\sc i\kern-.025em b}\kern-.08em
    T\kern-.1667em\lower.7ex\hbox{E}\kern-.125emX}}
\begin{document}

\title{Hovering Over the Key to Text Input in XR
\thanks{
$^1$ Google, USA. $^2$  University of Birmingham, UK. $^3$ University of Copenhagen, Denmark. $^4$  Arizona State University, USA. $^5$  Northwestern University, USA

$*$ equal contribution
}
}
\author{

\IEEEauthorblockN{Mar Gonzalez-Franco* $^1$, Diar Abdlkarim* $^2$, Arpit Bhatia $^3$, Stuart Macgregor $^2$,
Jason Alexander Fotso-Puepi $^1$}

\IEEEauthorblockN{
Eric J Gonzalez $^1$,
Hasti Seifi $^4$,
Massimiliano Di Luca $^2$,  
Karan Ahuja $^{1,5}$}
}

\maketitle

\begin{abstract}

Virtual, Mixed, and Augmented Reality (XR) technologies hold immense potential for transforming productivity beyond PC. Therefore there is a critical need for improved text input solutions for XR. However, achieving efficient text input in these environments remains a significant challenge. This paper examines the current landscape of XR text input techniques, focusing on the importance of keyboards (both physical and virtual) as essential tools. We discuss the unique challenges and opportunities presented by XR, synthesizing key trends from existing solutions.

\end{abstract}


\section{Introduction}
Keyboards remain the primary tool for efficient text input across personal computers and mobile devices. However, achieving comparably efficient text entry in XR environments has proven to be a significant challenge. Existing solutions are either inefficient, have limited accuracy, or require cumbersome physical setups. Without proper text input methods in XR, the development of productivity tools, immersive \textit{metaverse} experiences, and potential \textit{killer} apps for super-productivity remains hindered \cite{gonzalez2024guidelines}. 


The unique challenges of XR environments necessitate tailored approaches. Technical constraints, such as high-resolution displays and accurate finger tracking, can impede traditional input methods. Interestingly, the term "Metaverse" – popularized by Neal Stephenson's "Snow Crash" – has an intriguing connection to keyboards. The Metaverse originally signified a space dominated by those proficient with the "Meta" key, a function key, like the modern Ctrl, Shift or Alt, found on early keyboards that first appeared on the Stanford Artificial Intelligence Lab (SAIL) keyboard in 1970 marked by a black diamond\footnote{http://en.wikipedia.org/w/index.php?title=Meta\%20key}. This historical link hints at the potential for innovative XR text entry methods to augment productivity in ways that extend beyond traditional written input and document creation. Maybe the real Meta-verse is just a keyboard-enabled XR.

This paper synthesizes aspects that have been consistently explored as solutions, and identified as challenges for this field. We do so by partially exploring the growing body of research that has focused on novel XR text input techniques.

\section{Design Considerations}
 From the very beginning the goal of keyboard design is to allow users to enter text into the computer system by pressing buttons as fast and accurately as possible by optimizing the spatial layout as well as using sounds and haptics from mechanical keys. In a way this is a legacy solution as Keyboards pre-date digital world, and go as far back as to mechanical typewriters. With modern sensing and ML capabilities text input has evolved significantly, and keyboards can be changed in shape and form and adapt to user needs \cite{findlater2012personalized}.  

A similar progress could be traced into the future of text input in XR. In their 1997 paper, Mine et al. noted the difficulty of text entry in virtual environments \cite{mine1997moving}. One of the earliest attempts to addressing this challenge was VType \cite{evans1999vtype} which leveraged a commercial glove for tracking all 10 fingers and mapped multiple characters to each finger together with statistical analysis to disambiguate characters. Twenty five years later, effective text entry remains an open challenge in XR for consumers. 

Currently average typists on PC using 10 fingers (touch typing) in PC can achieve 40-60 WPM, with peaks of 80 Words Per Minute (WPM), while hunt-and-peek typists will achieve 27-37 WPM \cite{feit2016we}. 

In fact, there are several key metrics and considerations that come into play when it comes to quantitative metrics beyond WPM, like the N-Key Rollover (independent recognition of simultaneous keypresses), throughput, correctable typing error (percentage of word errors corrected by a language model, and character edit distance for correction accuracy). It is also important to track other typing errors like incorrect key registrations, multiple registrations from a single press, and missed key presses. As well as subjective metrics like the NASA Task Load Index that can offer insights into users' cognitive workload. When WPM are very low users tend to utilize dictation via voice-to-text \cite{ruan2018comparing}, despite voice's privacy and throughput limitations.

Across the board, XR text input performs worse than PC, virtual keyboards easily drop to 5 to 10 WPM, and even when using physical keyboards users can only keep 60
\% of their typing speed and 80\% of their accuracy even in VR \cite{dube2019text}. And the reason could be that in VR the technical requirements for text entry are stringent, especially for display resolution, FOV and finger tracking. In fact the complex tech stack needed for XR necessitates of particular design considerations. 

\subsection{Display Resolution}
Until very recently it was hard to read any text on commercial HMDs, much less to write. 
If we consider in 2017 VR displays usually had resolutions of 500-600 pixels per inch (PPI). As of 2023 most consumer displays had achieved 1200 PPI, despite LCDs in laboratories achieved 2000 PPI \cite{wu2023breaking}. Large enough MicroOLED displays with higher PPI numbers have only appeared as of 2024 with the Apple Vision Pro, that has 3386 PPI. And as of May 2024 LG has presented similar advances in MicroOLED with up to 4000 PPIs.

\subsection{Field of View} \label{sec:FOV}

While XR headsets often advertise wide fields of view or FOV (e.g., $110^{\circ}$H $\times$ $96^{\circ}$V), they still fall short when compared to the full extent of human vision ($200^{\circ}$H $\times$ $135^{\circ}$V). While human peripheral view often captures the user's own body, this isn't available in XR, while large FOV would be especially valuable for embodied interactions in XR \cite{nakano2021head}. 

\begin{figure}[h!]
    \centering
    \includegraphics[width=\columnwidth]{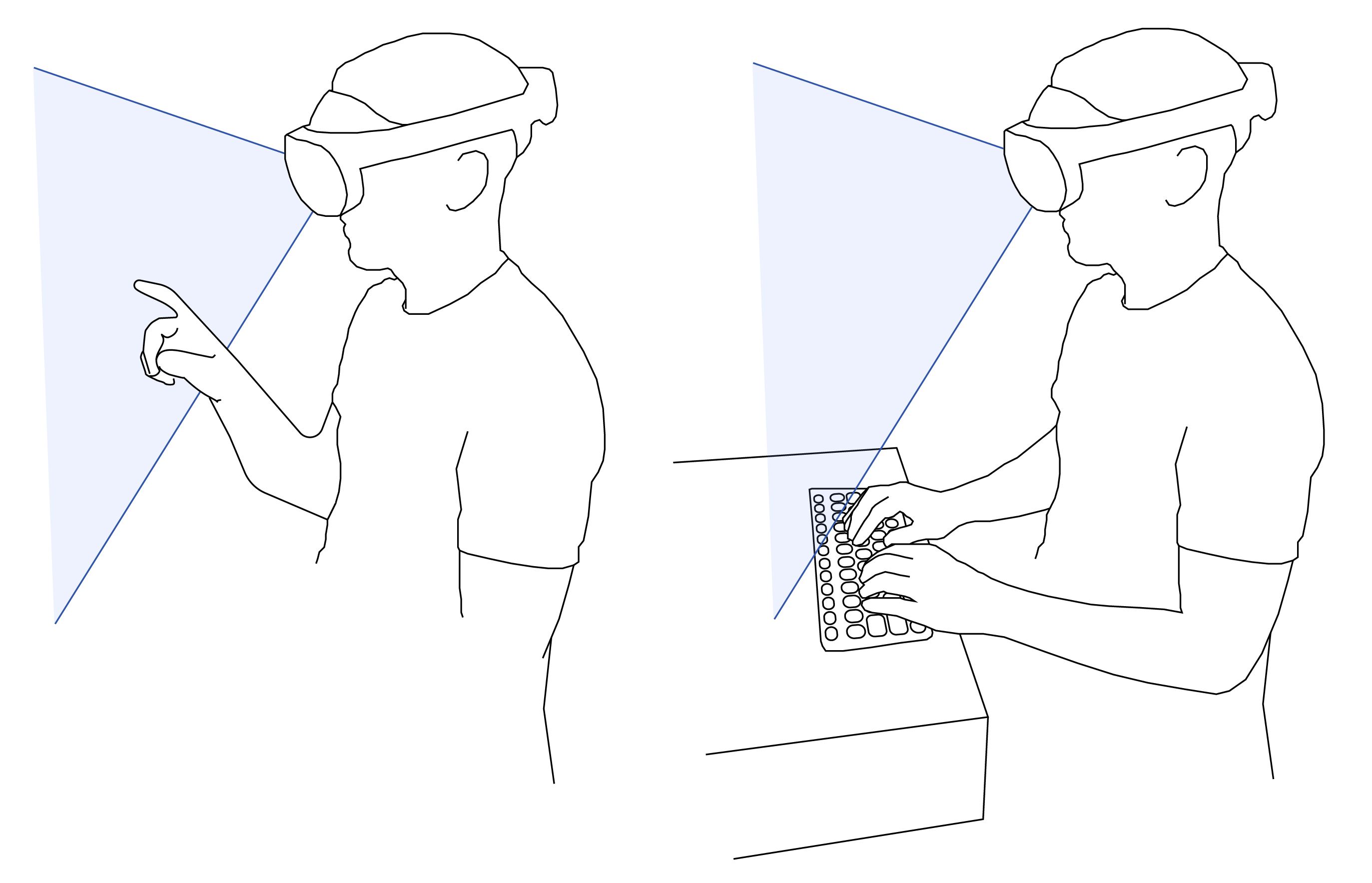}
    \caption{Ergonomic Issues of mid air touch include lack of precision and ergonomic issues with physical surface include field of view limitations. And additionally challenges on hand tracking from an egocentric view. As the finger tips are easily occluded by the rest of the hand. }
    \label{fig:ErgonomicIssues}
\end{figure}

When using a physical keyboard in XR (either in passthrough or with a virtual proxy), the limited vertical FOV of current HMDs means that users often cannot see the both the keyboard and the virtual content at the same time (Figure \ref{fig:ErgonomicIssues}). Furthermore, the lightseals of most HMDs also prevent the viewing of one's hands/keyboard by glancing below the display. Overall, this lack of ``glanceablility'' for physical keyboards in XR reduces ergonomics, and potentially hinders productivity.

\subsection{Finger Tracking and Typing}
While virtual keyboards offer a promising solution to this, the tracking of fingers and surfaces from the oblique, occlusion-prone, egocentric views provided by headset-mounted cameras remains technically challenging \cite{gu2020qwertyring}. Finger tracking often fails from a egocentric perspective (with the camera mounted on the head) precisely at the moment of surface touch, a problem that is difficult to solve due to self-occlusions, where the back of the hand occludes the fingers at certain hand orientations. And despite the fact that many people are working on hacks for text entry in VR on surfaces, it is still an unsolved problem \cite{gu2020qwertyring,liang2023shadowtouch}.


And without great finger tracking we end up with typing solutions that often rely in only two fingers (`hunt-and-peck'), which by all means is slower and less accurate than using all ten fingers \cite{rieger2012motor,feit2016we}. 

The hunt-and-peck method involves searching for keys with one finger from each hand, often looking at the keyboard instead of the content that’s being written. While seemingly easier at first, it is ultimately inefficient and dramatically limits typing speed.

In contrast, `touch typing' involves using all ten fingers, each assigned specific keys and resting on a "home row" \cite{rieger2012motor}. Even when not an expert touch typist, using more fingers and relying less on looking at the keyboard allows one to focus on the content, reduce exertion, and type faster with fewer errors \cite{feit2016we}. 

Up to now, the majority of text input methods for XR have fallen under the category of hunt-and-peck, either using direct touch with two index fingers or indirect raycasting from hands or controllers \cite{speicher2018selection}. Expanding to support multi-finger text input has the potential to greatly increase throughput, though technical hurdles such as handling self-occlusion and better finger accuracy during hand tracking must still be resolved. Furthermore, we expect that the benefits of ten-finger typing may be more evident in certain scenarios, such as when typing on a surface-aligned keyboard.



\subsection{Hand Representations}

While expert touch typists may rely less on visual cues, the majority of typists (even if proficient) tend to look at their keyboard at least occasionally during productivity sessions \cite{pinet2022typing}. Furthermore, inexperienced typists using the `hunt-and-peck' method heavily rely on visual feedback of both their hands and keyboard. The process of mastering a keyboard necessitates acquiring sensorimotor expertise through the integration of multiple sensory inputs. Visualizing the keyboard aids in learning the spatial arrangement of keys, while seeing the hands helps guide and maintain proper finger placement \cite{grubert2018effects}. Tactile feedback from the keys confirms successful presses and reinforces alignment, while auditory clicks offer additional confirmation \cite{ma2015haptic}. Finally, the visualized text output provides feedback on typing accuracy, closing the feedback loop \cite{feit2016we}.

Research suggests \cite{grubert2018effects} that experienced typists can achieve even faster or comparable typing speeds (WPM) on a physical keyboard when hands are represented abstractly, reducing visual occlusion. However, a significant drop in productivity and increased mental workload occurs when hand tracking is unavailable \cite{grubert2018effects}. This highlights the critical role hand representations play in XR keyboard interactions \cite{canales2019virtual}, especially for novice or less-skilled typists.  

\subsection{Flexible Keyboard Layouts}
XR presents a unique opportunity to reimagine how we view and interact with our keyboard. With XR, it is possible to scale, duplicate, offset, or otherwise augment the displayed keyboard. We believe this presents an opportunity to potentially enhance the text entry experience.

For example: as mentioned in Section \ref{sec:FOV}, typing on a physical keyboard in limited-FOV XR creates an ergonomic problem because users must choose between looking down at their keyboard or up at the content. One potential solution is to duplicate keyboards:
one anchored to a physical keyboard (or tabletop) for tactile feedback and optimal hand positioning, and a second visually re-projected keyboard positioned closer to the content. 
Fostering more robust motor control through cross-modal binding, crucial to typing which is a motor control task. This approach may align with active inference frameworks, potentially minimizing prediction errors for a smoother typing experience \cite{maselli2022active}. 


\subsection{Other considerations}
Additionally, proprioception, task agency, and embodiment significantly impact the keyboard experience. Social acceptability (from both first-person and observer perspectives) should be considered as well. Finally, other design considerations such as ergonomic factors are crucial, including mental and physical fatigue, haptic feedback, and device availability.

\section{Opportunities for Keyboards in XR}



\subsection{The Case for a Physical Keyboard}

Physical keyboards offer distinct haptic feedback, familiar layouts, and the potential for high throughput text input. In XR, keyboard representations often rely on either 3D tracking (e.g., Quest \cite{abdlkarim2022methodological}) or passthrough visualization \cite{gruen2020measuring}.
However physical keyboards might not always be available and they also suffer from some of the design considerations mentioned before such as FOV, hand representations and passthrough needs.


\subsection{The Need for a Virtual Keyboard}

Virtual keyboards 
eliminate the need for external hardware and offer greater portability. However, virtual keyboards are associated with  slow speeds. 

\subsubsection{Mid-air Virtual Keyboards}

Mid-air virtual keyboards are a conceptually intuitive and a widely explored approach for text input in XR. They often mimic traditional keyboard layouts (e.g., QWERTY) and are positioned within the user's field of view. Input typically involves either hand \cite{benoit2017bimanual}, controller-based pointing \cite{gupta2023investigating } or the use of eye gaze \cite{gupta2023investigating,Pfeuffer2024gaze} for key selection. The appeal of this approach lies in its relative simplicity, potential for leveraging existing user familiarity, and minimal hardware requirements beyond standard XR devices.

However, mid-air virtual keyboards also face significant challenges. The absence of physical surfaces and haptic feedback makes precise key targeting difficult, leading to higher error rates \cite{gupta2020tactilemidair}. Extended use frequently causes fatigue and discomfort in the arms and shoulders, particularly with hand-based input \cite{cheng2022comfortable}. This lack of physicality and the reliance on `hunt-and-peck' style interactions severely limit typing speed compared to traditional keyboards \cite{mcgill2015dose}. Additionally, mid-air keyboards can obstruct the virtual environment, potentially hindering immersion and task performance.

however it is also possible we see a shift in the way users interact with mid-air keyboards that could involve methods inspired by swipe-based smartphone typing \cite{boustila2019text}, like the work by Dudley et al. \cite{dudley2023evaluating}, novel ergonomic raycast techniques tailored to XR akin to the CD gain of a mouse \cite{kumar2020tagswipe}, or even entirely new interaction modalities that capitalize on the unique sensing and ML capabilities of future XR devices.


\subsubsection{Surface-anchored Virtual Keyboards}

A growing trend in XR text input is the use of surface-anchored virtual keyboards. The advantages of leveraging surfaces are mostly supporting of ergonomics and providing additional haptics \cite{cheng2022comfortable}. Surfaces aren't just tables, and can range from dedicated wearables like watches \cite{xia2015nanostylus} to the user's own body, including fingernails \cite{chan2013fingerpad} or finger tips \cite{xu2019tiptext}. By making use of a physical surface, these techniques offer some inherent haptic feedback, enhancing accuracy and reducing fatigue compared to mid-air keyboards \cite{xia2015nanostylus}. The potential for compact, faster and even discreet text entry makes surface-anchored solutions appealing in situations where traditional keyboards are impractical.


\subsubsection{ML-enabled Keyboards}

In virtual keyboards, "tap detection" identifies individual key selections by the user. 
However, raw tap detection has proved challenging for robust text entry.
Hence XR keyboards relying solely on tap detection would be slow and prone to errors.

Machine learning (ML) decoding models offer a solution by analyzing sequences of taps rather than individual inputs. These models consider statistical patterns of language and user input behavior, enabling them to correct likely typos, predict words, and personalize suggestions \cite{alharbi2020effects,zhai2012word}. This significantly enhances accuracy and speeds up text entry, mirroring the transformative impact of predictive models such as those used by Gboard  for smartphone typing.

Probabilistic language models (PLMs) like Bayesian Neural Networks (BNNs) are particularly well-suited for this task \cite{streli2022taptype}. Their ability to manage uncertainty in tap data is crucial in real-world XR scenarios where environmental factors and varying user behavior might lead to noisy or ambiguous input. By distributing probabilities across potential interpretations, BNNs increase the accuracy of the text decoding process.

Beyond PLMs, other ML approaches hold promise. Recurrent Neural Networks (RNNs) are adept at analyzing sequential data and capturing dependencies between taps \cite{mrazek2021using}.  When combined with Convolutional Neural Networks (CNNs), they can analyze both temporal and spatial patterns of taps \cite{meier2021tapid}, potentially enhancing accuracy in scenarios where precise tap location is informative (e.g., surface-based keyboards).  As AI and large language models advance, their integration with XR text input becomes increasingly compelling.  Techniques like "sensor tap-to-language token embedding" \cite{jin2024position} could bridge the gap between raw input and sophisticated language models, leading to further breakthroughs in  intuitive and efficient XR text input experiences.

\section{Discussion and Conclusions}

Good UI and interaction tools help us learn the technology limits. A rare function key beyond Crtl+C, is currently on PC very hard to discover and or learn. But a Keyboard inside XR, is in the end a software designed UI, that can facilitate that learning through contextual augmentation. 
The learning might then surprisingly transfer back to PC and other less immersive devices. So we are not just augmenting the keyboard but augmenting the human. 

The diverse use cases of XR, ranging from office work, to gaming, entertainment, to on-the-go applications, paired with the rapid advances of XR technologies introduce new interaction possibilities and challenges for text input.  



This paper has showcased the main trends and challenges we have observed when exploring the field of text input and productivity in XR. 

Experts on the real world unequivocally recommend learning touch typing. Although investing in touch typing might seem like an extra effort, it is a valuable skill that benefits in the long run, saving time, improving accuracy, and boosting overall productivity. But when it comes to XR typing there isn't a clear recommendation, with many choices.


Perhaps the solution to text entry is indeed not a single one, as we might actually need a solution for physical keyboard, another one for virtual keyboards, depending on when surfaces are available and when they are not, it also includes improving ML algorithms, and optimizing typing assisted by existing devices phones or watches. 


\bibliographystyle{IEEEtran}

\bibliography{template}

\end{document}